\documentclass[12pt,preprint]{aastex}

\lefthead{Bekki et al.}
\righthead{Origin of  globular cluster systems}

\begin{document}

\title{
The U-shaped distribution of globular cluster specific frequencies
in a biased globular cluster  formation scenario}

\author{Kenji Bekki} 
\affil{
School of Physics, University of New South Wales, Sydney NSW 2052, Australia}

\author{Hideki Yahagi}
\affil{
Department of Astronomy, University of Tokyo, 7-3-1 Hongo, Bunkyo
ward, Tokyo 113-0033, Japan\\}

\and

\author{Duncan  A. Forbes}
\affil{Centre for Astrophysics \& Supercomputing,
Swinburne University of Technology,
Hawthorn, VIC 3122, Australia,
\\}

\begin{abstract}

Using high-resolution numerical  simulations,
we investigate mass- and luminosity-normalized specific frequencies 
($T_{\rm N}$ and $S_{\rm N}$, respectively) of globular cluster systems
(GCSs) in order to understand the origin of the observed
U-shaped relation between 
$S_{\rm N}$ and  $V$-band magnitude ($M_{\rm V}$) of their host galaxies.
 We adopt a biased GC formation scenario in which 
GC formation is truncated in galaxy halos that are virialized at
a later redshift, $z_{\rm trun}$. $T_{\rm N}$ is derived for 
galaxies with GCs today and converted into $S_{\rm N}$ for reasonable
galaxy mass-to-light-ratios ($M/L$).
We find that $T_{\rm N}$ depends on halo mass ($M_{\rm h}$)
in the sense that  $T_{\rm N}$ can be larger in more massive
halos with $M_{\rm h} > 10^9 {\rm M}_{\odot}$,
if $z_{\rm trun}$ is as high as 15.
We however find that the dependence is  too weak to explain the observed
$S_{\rm N}$-$M_{\rm V}$ relation and the wide range of $S_{\rm N}$
in low-mass early-type galaxies with $-20.5 < M_{\rm V}  < -16.0$  mag
for a reasonable constant $M/L$.
The $M_{\rm V}$-dependence  
of $S_{\rm N}$ 
for the low-mass galaxies
can be well reproduced,
if the mass-to-light-ratio 
$M_{\rm h}/L_{\rm V} \propto {M_{\rm h}}^{\alpha}$,
where $\alpha$ is as steep as $-1$.
Based on these results, we propose that
the origin of the observed U-shaped $S_{\rm N}$-$M_{\rm V}$
relation of GCSs
can be understood in terms of the bimodality in
the dependence of $M_{\rm h}/L_{\rm V}$ on $M_{\rm h}$
of their host galaxies.
We also  suggest that the observed large dispersion
in $S_{\rm N}$ in low-mass galaxies is due partly
to the large dispersion in $T_{\rm N}$.

\end{abstract}

\keywords{
globular clusters: general --
galaxies: star clusters --
galaxies: evolution -- 
galaxies: stellar content
}

\section{Introduction}

Specific frequencies ($S_{\rm N}$) of globular clusters (GCs) in 
galaxies are observed to be different between galaxies
with different Hubble morphological types
and between those in different environments 
(e.g., Harris 1991).
The $S_{\rm N}$ of globular cluster systems (GCSs)  
and  their correlations with  physical  properties 
of their host galaxies 
have long been discussed in various
different contexts of galaxy formation; biased GC formation
in high-density environments (e.g., West 1993), 
origin of field and cluster elliptical galaxies 
(e.g., Ashman \& Zepf 1992; Forbes et al. 1997),
and origin of dwarf elliptical (dE) galaxies (e.g., Miller et al. 1998).
Recent numerical and theoretical studies have also discussed the origin of
$S_{\rm N}$ of elliptical galaxies interacting with
cluster environments (e.g., Bekki et al. 2003).

Durrell et al. (1996) investigated the trend between
$S_{\rm N}$ and $M_{\rm V}$ (where $M_{\rm V}$ is the $V$-band
magnitude of a galaxy) for galaxies with a wide range of $M_{\rm
V}$. They found that $S_{\rm N}$ values were higher 
in the lowest luminosity dwarf ellipticals (dEs), and similar to the $S_{\rm N}$
values for giant ellipticals (see Fig. 9 in their paper). $S_{\rm
N}$ values appeared to have a minimum at intermediate galaxy
luminosities, but there were very few data points available to
them. Forbes (2005) recently investigated the 
$S_{\rm N}$ values for $\sim$100 early-type galaxies in the Virgo
cluster, confirming the high $S_{\rm N}$ values for dwarfs and
low $S_{\rm N}$ values at $M_{\rm V}$ $\sim$ --19.5. 
Thus the $S_{\rm N}$-$M_{\rm V}$ relation reveals a `U'-shaped,
or bimodal, distribution.
Durrell et al. (1996) speculated that
supernova-driven mass loss was responsible for the higher $S_{\rm
N}$ values in dwarfs. Forbes (2005) suggested that this, and the
transition from hot to cold gas accretion (Dekel \& Birnboim
2005) played a role in
determining the number of GCs per unit galaxy starlight. He
further noted that this increase in  $S_{\rm N}$ closely followed
the dependence of galaxy $M/L$ with mass.  
However, this observed $S_{\rm
N}$ bimodality
has not yet been explored by theoretical/numerical studies of
GC formation.

The purpose of this {\it Letter}  is to demonstrate, for the first time,
whether and how  the observed  U-shaped  
$S_{\rm N}$-$M_{\rm V}$ relation is reproduced in 
a ``biased GC formation  scenario'' in which 
GCs can be formed only in galaxy subhalos with high 
primordial densities and thus virialized earlier
(i.e., further GC formation is truncated after a redshift
of $z_{\rm trun}$).
This biased GC formation scenario 
has been discussed in the context of
GCS properties    (e.g., West 1993; Beasley et al. 2002;
Santos 2003; Bekki 2005; Yahagi \& Bekki 2005; Moore et al. 2005),
but not discussed in the context of 
the U-shaped $S_{\rm N}$ distribution.
Throughout this paper,  we adopt the standard
definition of 
$S_{\rm N}=N_{\rm GC}\times 10^{-0.4(M_{\rm V}+15)}$
and $T_{\rm N}=T_0 N_{\rm GC}/M$, where $N_{\rm GC}$ is
the total number of GCs in a GCS and
$T_0$ is a normalization factor such that
$T_{\rm N}$ has reasonable values (an order of $1-10$).

\section{The observed U-shaped relation  and its fitting function}

Figure 1 shows the observed U-shaped  
$S_{\rm N}$-$M_{\rm V}$ relation from
Harris (1991) for bright galaxies and from Durrell et al. (1996)
for dwarf ones. 
It is clear from this figure that the galaxy luminosity dependence
of $S_{\rm N}$ is different between galaxies
below and above a threshold luminosity, $M_{\rm V,th}$.
Although the dependence is less clear for galaxies
with $M_{\rm V} < M_{\rm V,th}$ in Figure 1,
Rhode et al. (2005)  found a strong trend of increasing
$T_{\rm N}$ with increasing galaxy luminosity for elliptical galaxies
with mass $> 10^{11} {\rm M}_{\odot}$.

The observed $S_{\rm N}$-$M_{\rm V}$ and $T_{\rm N}$-$M$
relations may provide vital clues to galaxy and GC
formation (e.g., Harris 1991; Forbes 2005).
Accordingly  comparison between observed and simulated relations
may well enable us to derive some physical meaning for 
the relations. 
It is therefore important to quantify the relations by using
an admittedly simple mathematical expression.
We here propose the following form for the $S_{\rm N}$-$M_{\rm V}$
relation; 
\begin{equation}
S_{\rm N} (x) = A_1 \times 10^{K_1 x}+ (S_{\rm N,th}-A_1) \times
10^{K_2 x},
\end{equation}
where $x=\frac{M_{\rm V}-M_{\rm V,th}}{M_{\rm V,th}}$ and
$S_{\rm N,th}$ is  $S_{\rm N}$ at $M_{\rm V}=M_{\rm V,th}$.
Parameter values of $A_1$, $K_1$, $K_2$, $M_{\rm V,th}$,
and $S_{\rm N,th}$ can be determined by fitting to observations.
An example of the model fit with
$M_{\rm V,th}=-19.5$ mag,
$S_{\rm N,th}=1.0$, $A_1=0.5$,  $K_1=-6$, and  $K_2=4$ 
is shown in Figure 1.

This functional form in equation (1) can be equivalent to
$B_1 \times {L_{\rm V}}^{C_1} + B_2 \times {L_{\rm V}}^{C_2}$,
where $L_{\rm V}$ is the $V$-band luminosity and other parameter
values (e.g., $B_1$) can be derived from values of the parameters 
(e.g., $A_1$) in equation  (1). Therefore, the functional form
has some advantages in discussing 
both theoretically and observationally the origin of $S_{\rm N}$ 
(and $T_{\rm N}$) in terms of mass and luminosity growth
processes of galaxy assembly. 
The observed $\gamma$ of $1-2$ 
in $N_{\rm GC} \propto {L_{\rm V}}^{\gamma}$ for bright ellipticals 
(Zepf \& Ashman 1993) suggests that  $C_2$ can range from 0 to 1.

 
\section{Numerical simulations of $T_{\rm N}$ and $S_{\rm N}$ }

\subsection{The model}

Since methods and techniques of
numerical simulations of GC formation at high redshifts
are given in detail in our previous paper (Yahagi \& Bekki 2005, YB),
we only briefly describe them in this paper.
We simulate the large scale structure of GCs  
in a $\Lambda$CDM Universe with ${\Omega} =0.3$, 
$\lambda=0.7$, $H_{0}=70$ km $\rm s^{-1}$ ${\rm Mpc}^{-1}$,
and ${\sigma}_{8}=0.9$ 
by using the Adaptive Mesh Refinement $N-$body code developed
by Yahagi (2005) and Yahagi \& Yoshii  (2001).
We use $512^3$ collisionless dark matter (DM) particles in a simulation
with the box size of $8.26h^{-1}$Mpc and a particle mass 
of $5 \times 10^{5} {\rm M}_{\odot}$. 
We start simulations at $z=41$ and follow them until $z=0$
in order to investigate the physical properties
of old GCs outside and inside virialized dark matter halos at $z=0$. 
We used the COSMICS (Cosmological Initial Conditions and
Microwave Anisotropy Codes), which is a package
of fortran programs for generating Gaussian random initial
conditions for nonlinear structure formation simulations
(Bertschinger 1995, 2001). 

We  assume that (1) old, metal-poor globular cluster (MPC) formation
is truncated for $z \le z_{\rm trun}$,
and (2)  initial radial  profiles of GCSs in
virialized subhalos at $z = z_{\rm trun}$
are the same as those of the
``NFW'' profiles
(Navarro, Frenk, \& White 1996).
We follow dynamical evolution of GCs 
from $z = z_{\rm trun}$ to 0 and
thereby investigate mass of the virialized halos at $z=0$
($M_{\rm h}$), the total numbers of GC particles  within the halos
($N_{\rm GC}$),
and  the normalized number of GCs per halo
mass ($T_{\rm N}=T_0 N_{\rm GC}/M_{\rm h}$).
Given the fact that there are currently no observational
constraints on initial GCS density profiles at  $z = z_{\rm trun}$,
we consider that adopting the NFW profiles
is a reasonable first step for better understanding GCS properties
in the CDM model.

We convert the simulated $T_{\rm N}$-$M_{\rm h}$ relations
into  the $S_{\rm N}$-$M_{\rm V}$ ones by using the following formula:
\begin{equation}
S_{\rm N} =  S_0 \frac{N_{\rm GC}}{L_{\rm V}} =
S_0  \left(\frac{N_{\rm GC}}{M_{\rm h}}\right)
\left(\frac{M_{\rm h}}{L_{\rm V}}\right)
= S_0  \left(\frac{T_N}{T_0}\right)\left(\frac{M_h}{L_V}\right)
\end{equation}
where $S_0$ is a constant
and the normalization  factor $T_0$ for each model
is determined such that number of GCs 
per unit mass in  a simulation box is the same between models
with different $z_{\rm trun}$.
We then compare the simulated relations 
to the observed data of Forbes (2005) in which
the $S_{\rm N}$ of blue GCs was derived based on 
new data from the ACS Virgo Cluster Survey (Peng et al. 2005).

We investigate two $M/L$ models, i.e. variable and constant  $M/L$.
For the variable $M/L$ model, 
we assume that 
$M_{\rm h}/L_{\rm V} \propto {M_{\rm h}}^{\alpha}$
for $M_{\rm h} \le M_{\rm h,th}$,
where $M_{\rm h,th}$ is set to be $10^{11} {\rm M}_{\odot}$,  and
$M_{\rm h}/L_{\rm V} \propto {M_{\rm h}}^{\beta}$
for $M_{\rm h} > M_{\rm h,th}$.
We here focus mainly on low-luminosity galaxies for which observational
data of $S_{\rm N}$ is available (Forbes 2005).
We accordingly chose 
the value of $\beta$ 
consistent with observations based on galaxy luminosity functions
(i.e. $\beta=0.33$;  Marinoni \& Hudson 2002)
and derive the best value of ${\alpha}$ for low-luminosity galaxies.
Although we investigate models with a wide range
of  $M_{\rm h}/L_{\rm V}$ (i.e. $5-100$) for the constant $M/L$ model, 
we only show the results of the model with $M_{\rm h}/L_{\rm V}=10$.
It should be stressed here that all of the present results
are galaxies containing GCs at $z=0$; GC-less galaxies 
at $z=0$ (typically those with $M_{\rm V} > -15$) 
are not included in the present study.

Although Forbes (2005) speculated about  a possible $M$-dependent $M/L$
explaining the observed $S_{\rm N}$-$M_{\rm V}$ relation,
his adopted assumptions (e.g., constant $T_{\rm N}$) are 
not based on detailed theoretical/numerical studies.
Since the present study derives $S_{\rm N}$ from 
the simulated $T_{\rm N}$,  we can discuss 
more self-consistently the origin
of the $S_{\rm N}$-$M_{\rm V}$ 
relation on firm physical basis. 
We demonstrate that the simulated $T_{\rm N}$-$M_{\rm h}$
dependences are too weak to explain the wide range of $S_{\rm N}$
for a constant $M/L$:
The difference in GC number per unit galaxy mass between 
galaxies alone  can not
explain observations.

\subsection{Results}

Figure 2 shows the distributions of the simulated GCSs at $z=0$ 
in the $T_{\rm N}$-$M_{\rm h}$ plane 
for the models with $z_{\rm trun}=6$ and 15.
Below a critical halo mass
of $\sim  2 \times  10^{9} {\rm M}_{\odot}$, 
$T_{\rm N}$ is higher for halos with smaller  $M_{\rm h}$
in the model with $z_{\rm trun}=15$. 
Above the critical mass, $T_{\rm N}$ is higher for
halos with larger $M_{\rm h}$ in this model
and the slope of
the $M_{\rm h}$-dependence of  $T_{\rm N}$ is
significantly shallower than in halos below $M_{\rm h, th}$.
This clear bimodality in  the $T_{\rm N}$-$M_{\rm h}$ relation
is not seen in the model with  $z_{\rm trun}=6$,
which shows a monotonically increasing trend of $T_{\rm N}$ with
$M_{\rm h}$ for 
$10^8 {\rm M}_{\odot} \le M_{\rm h} \le 3 \times 10^{10} {\rm M}_{\odot}$.
The dispersion in $T_{\rm N}$ appears to be 
significantly larger in halos with smaller $M_{\rm h}$
for both models, which implies that
the observed large dispersion in $S_{\rm N}$
in low-mass galaxies can be understood in terms of
the large dispersion in $T_{\rm N}$.

For the model with $z_{\rm trun}=15$,
GC formation is assumed to occur only in the subgalactic
halos formed from rare high-density peaks in the primordial matter
distribution
and thus virialized before $z=15$.
Accordingly, the bimodality 
can be understood in terms of the following two competing effects
of halo merging with, or without, GCs.
Low-mass subhalos with initially high $T_{\rm N}$
(virialized at $z>15$) grow  after their formation by merging/accretion
of subhalos most of which  are virialized later than $z=15$ and thus
are those without GCs (referred to as ``GC-less'' halos). 
Consequently, their $T_{\rm N}$ values 
can decrease as their masses become larger
during hierarchical growth.
{\it This is the first effect
of halo merging.}

According to the theory of biased galaxy formation (e.g., Kaiser 1984),
the primordial density field of a more massive  halo today is
more likely to contain a larger fraction of high-density peaks.
Accordingly, a larger halo at $z=0$ can be formed
from a larger fraction of subhalos that are virialized before $z=15$
and which contain GCs (``GC'' halos).
Therefore,  a larger halo is more likely to be formed from
merging of a larger fraction of GC halos and thus  merging can
increase  $T_{\rm N}$ for larger halos. {\it This is the second effect
of halo merging. }

As a result of these two competing effects,
we can understand the bimodal $T_{\rm N}$-$M_{\rm h}$ relation,
i.e. below $M_{\rm h,th}$ the first effect is stronger, while
above the second effect is stronger.
The dramatic effect of the first GC-less halo merging is not 
expected for the model with $z_{\rm trun}=6$,
because most of subgalactic halos that are building blocks
of larger halos at $z=0$ have been virialized before  $z=6$
and contain GCs. 
Therefore the monotonically increasing
$T_{\rm N}$-$M_{\rm h}$ relation is 
due to the second effect in this model.
It should be stressed here that 
irrespective of $z_{\rm trun}$, $T_{\rm N}$ is more likely
to be larger for halos 
with $M_{\rm h} > 10^9 {\rm M}_{\odot}$.

Figure 3 shows the comparison between the observations (Forbes 2005)
and the best variable $M/L$ model with $\alpha \approx -1$,
where $\alpha$ is the slope  
in the relation of $M_{\rm h}/L_{\rm V} \propto {M_{\rm h}}^{\alpha}$. 
It should be noted here  that the observed $S_{\rm N}$ of blue GCs
in ellipticals brighter than $M_{\rm V} \sim -20.5$ mag 
are seriously underestimated due to the limited ACS field-of-view
(Forbes 2005). Thus the comparison
makes sense only for lower luminosity galaxies.
As shown in Figure 3, 
both models with $z_{\rm trun}=6$ and $z_{\rm trun}=15$
can reproduce well the observed trend of increasing $S_{\rm N}$ with
decreasing luminosity for low-mass galaxies with
$M_{\rm V} > M_{\rm h,th}$.

Figure 4 shows that the simulated $S_{\rm N}$-$M_{\rm V}$ relations
do not match well with the observations 
for the constant $M/L$ models.
We also find that
constant  $M_{\rm h}/L_{\rm V}$ models, with a wide range of
$M_{\rm h}/L_{\rm V}$  values (i.e. $5-100$), do not match with the
observations either.
These results, combined with those in
Figure 3, imply that the origin of the 
$S_{\rm N}$-$M_{\rm h}$ relation is closely associated
with the dependence of $M_{\rm h}/L_{\rm V}$ on $M_{\rm h}$, as
suggested by Forbes (2005). 
Figure 4 shows that $S_{\rm N}$ can be only slightly higher 
in low-luminosity galaxies with $M_{\rm V} >  -16.5$ mag
as a result of biased GC formation in
the model with $z_{\rm trun}=15$. However the simulated
dependence is not strong enough to explain quantitatively the observation
for the constant $M_{\rm h}/L_{\rm V}$ model.
Since observations for 
galaxies brighter than  $M_{\rm V}=-16$ mag should be compared
with simulations,  the results 
for $M_{\rm h} > 2 \times 10^{9} {\rm M}_{\odot}$
in Figure 2 are relevant to Figure 4. 
The derived $S_{\rm N}$-$M_{\rm V}$ relations  in Figure 4
are thus essentially the same as
$T_{\rm N}$-$M_{\rm h}$ ones for $M_{\rm h} > 2 \times 10^{9} {\rm M}_{\odot}$
in Figure 2.
If we adopt a
higher M/L, the blue/red lines are shifted to the left in Figure  4:
A model with M/L = 25 results in a 1 mag shift to lower luminosities.


\section{Discussion and conclusions}

West (1993) first suggested that the observed dependence of $S_{\rm N}$
on galaxy type and environment can be understood in terms
of GC formation from  the statistical enhancement of rare high-density 
peaks in the primordial matter distribution (i.e., biased GC formation).
The present numerical simulations have shown 
that biased GC formation in subgalactic halos virialized early
($z \sim 15$), can be imprinted on the observed 
$T_{\rm N}-M_{\rm h}$ relation.
Semi-analytic galaxy formation models based on a
hierarchical clustering scenario
have also shown that models with a
truncation of GC formation at $z \sim 5$ 
can better reproduce the observed color bimodality of GCs in 
early-type galaxies (Beasley et al. 2002).
What physical mechanisms are responsible for the truncation
(or the severe suppression)  of GC formation in subgalactic halos
virialized at later times?

Santos (2003) first proposed that
cosmic reionization at high $z$ can suppress GC formation
and thus be an important factor for better understanding 
the origin of the physical properties of GCSs.
Recent numerical simulations on GCs and old stellar halos 
in galaxies have
confirmed that structural and kinematical properties of GCSs 
carry fossil information about the epoch of  
the reionization $z_{\rm reion}$ that can suppress GC formation 
(Bekki 2005; Bekki \& Chiba 2005; Moore et al. 2005; Rhode et al. 2005). 
Therefore, if $z_{\rm trun}$ = $z_{\rm reion}$,
the present results imply that
the $T_{\rm N}-M_{\rm h}$  relation
 can provide some constraints
on $z_{\rm reion}$.
Furthermore, we suggest that
these relations can be different
in different environments (or cosmic volumes investigated by
observations),
if reionization is spatially inhomogeneous 
(Benson et al. 2001; Wyithe \& Loeb 2004).  

We suggest that 
it is worthwhile for future observations 
to investigate the $T_{\rm N}$-$M_{\rm h}$
relation for  a statistically significant number of
galaxies, although it is a formidable task to derive total galactic
masses including dark matter halos. 
Such observational studies  can test biased GC formation at high redshifts
by comparing the results with the corresponding simulations.
The predicted $T_{\rm N}$ values change  only by a factor of $\sim 5$ for
a change in $M_{\rm h}$ by four orders of magnitude.
The observed difference in  $S_{\rm N}$ for
a $M_{\rm V}$ range of $\sim 10$ mag  (corresonding roughly
to a luminosity difference of factor $\sim 10^4$)
is  more than a factor of $\sim 30$ (see Figure 1).
Therefore the observed bimodal $S_{\rm N}$-$M_{\rm V}$ relation
can not be explained well by the present models without resorting
a mass-dependent $M/L$ (for which there is some
independent evidence). 
We accordingly suggest that the origin of the observed 
U-shaped $S_{\rm N}$-$M_{\rm V}$ relation
of GCSs can be
associated, at least partly,
with the bimodality in the mass-dependence 
of the $M/L$ ratio of their host galaxies.

\acknowledgments
We are  grateful to the anonymous referee for valuable comments,
which contribute to improve the present paper.
The numerical simulations reported here were carried out on 
Fujitsu-made vector parallel processors VPP5000
kindly made available by the Astronomical Data Analysis
Center (ADAC) at National Astronomical Observatory of Japan (NAOJ)
for our  research project why36b.
K.B. and D.A.F. acknowledge the financial support of the Australian Research 
Council throughout the course of this work.
H.Y. acknowledges the support of the research fellowships of the Japan
Society for the Promotion of Science for Young Scientists (17-10511).

\clearpage


\clearpage
\plotone{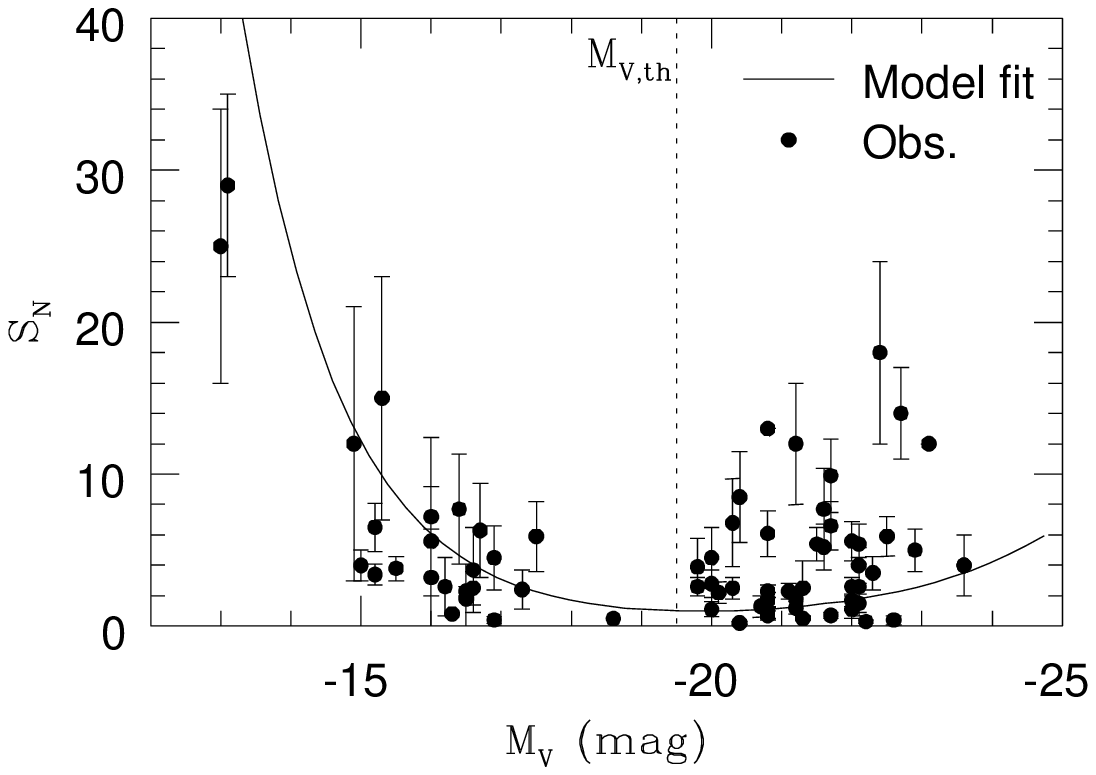}
\figcaption{
Observational data for GCSs in galaxies in the $S_{\rm N}$-$M_{\rm V}$
plane. Observational data are from Harris (1991) for bright galaxies
and from Durrell et al. (1996) for dwarf populations.
The solid line represents an empirical fit to the observed 
bimodality in the relation between $S_{\rm N}$ and $M_{\rm V}$
(the details of the functional form are given in the main text).
The dotted line marks the critical magnitude  ($M_{\rm V, th}$):
The $S_{\rm N}$ dependence on $M_{\rm V}$
is different between low and high luminosity galaxies. 
\label{fig-1}}

\clearpage

\plotone{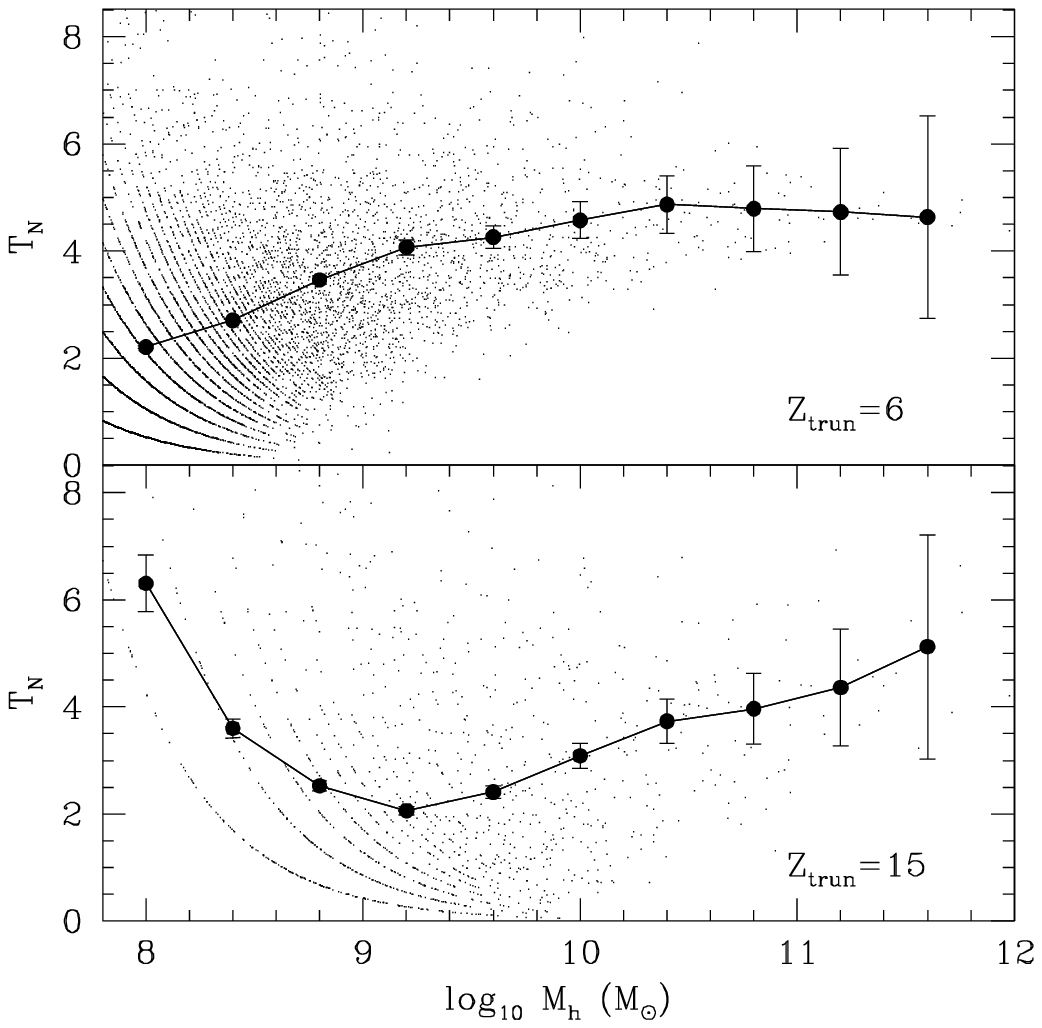}
\figcaption{
Dependences of $T_{\rm N}$ (=$T_0 N_{\rm GC}/M_{\rm h}$)
on ${\log}_{10} M_{\rm h}$ for the models
with $z_{\rm trun}=6$ (upper) and $z_{\rm trun}=15$ (lower).
Small dots represent simulated GCSs within galaxy-scale
halos of masses $M_{\rm h}$ and ``GC particle''
numbers  of $N_{\rm GC}$.
Big dots with error bars represent the mean values 
of  $T_{\rm N}$ in each mass bin. 
The error bar 
in each mass  bin is estimated as $T_{\rm N}/\sqrt{2(N_{i}-1)}$,
where $N_{i}$ is the total number of galaxy-scale halos in
the $i$-th mass bin.
In order to compare these results with observations,
the value of  the normalization factor ($T_{0}$) 
is determined for each model.
See text for details.
\label{fig-2}}

\clearpage
\plotone{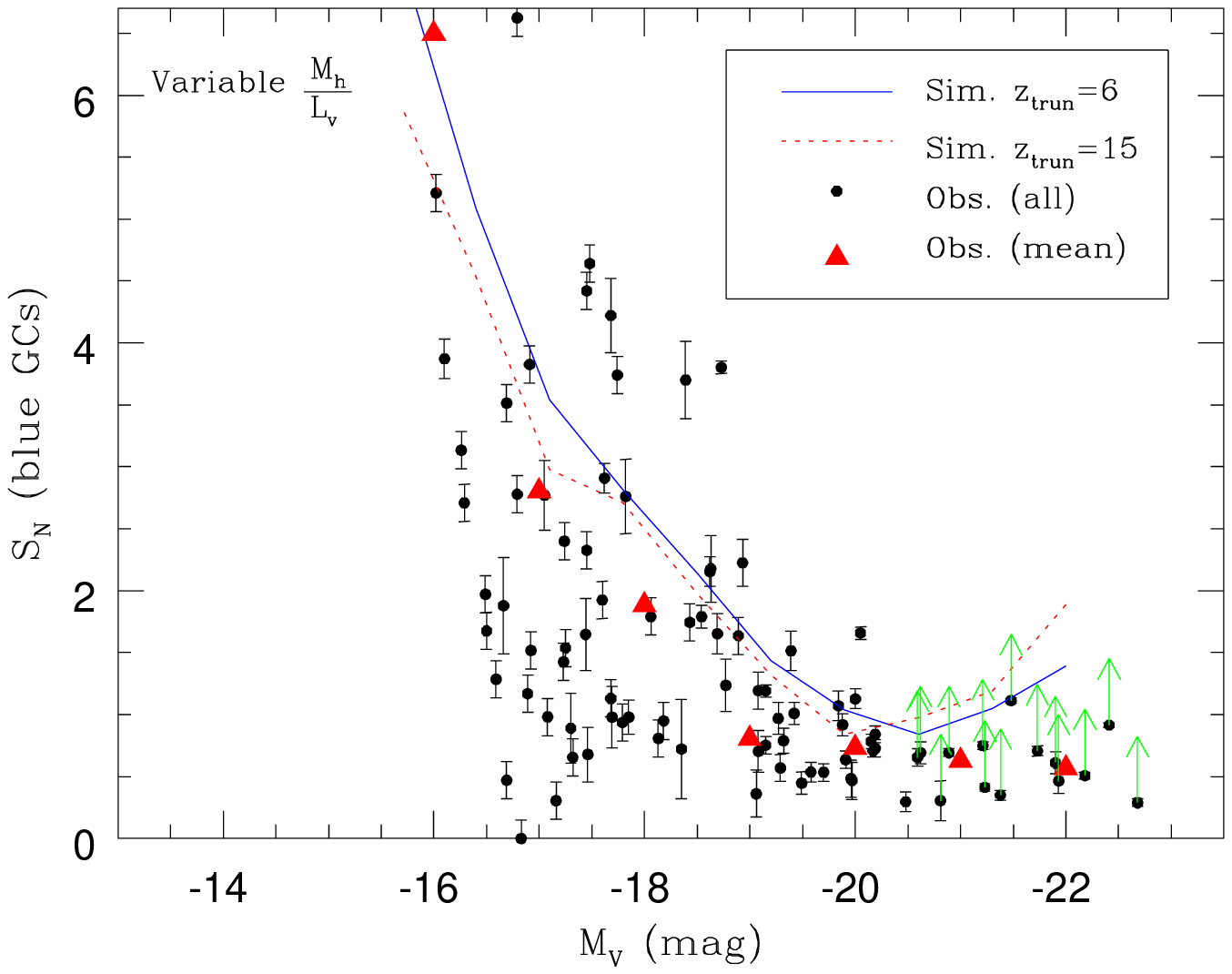}
\figcaption{
Observational data for $\sim$100 GCSs in early-type galaxies in
the $S_{\rm N}$-$M_{\rm V}$ plane. Here $S_{\rm N}$ 
values are shown for blue GCs only from Forbes (2005)
and the large triangles show the mean value of
$S_{\rm N}$ in each magnitude bin.
For comparison, 
the variable $M/L$ models with $M_{\rm h}/L_{\rm V} \propto {M_{\rm h}}^{-1.0}$
are shown for $z_{\rm trun}=6$
(solid) and $z_{\rm trun}=15$ (dotted).
Since $S_{\rm N}$ of brighter galaxies with $M_{\rm V}<-20.5$ mag
are significantly underestimated in observations,
the lower limits of $S_{\rm N}$ values are indicated by green arrows
for these galaxies.
\label{fig-3}}

\clearpage
\plotone{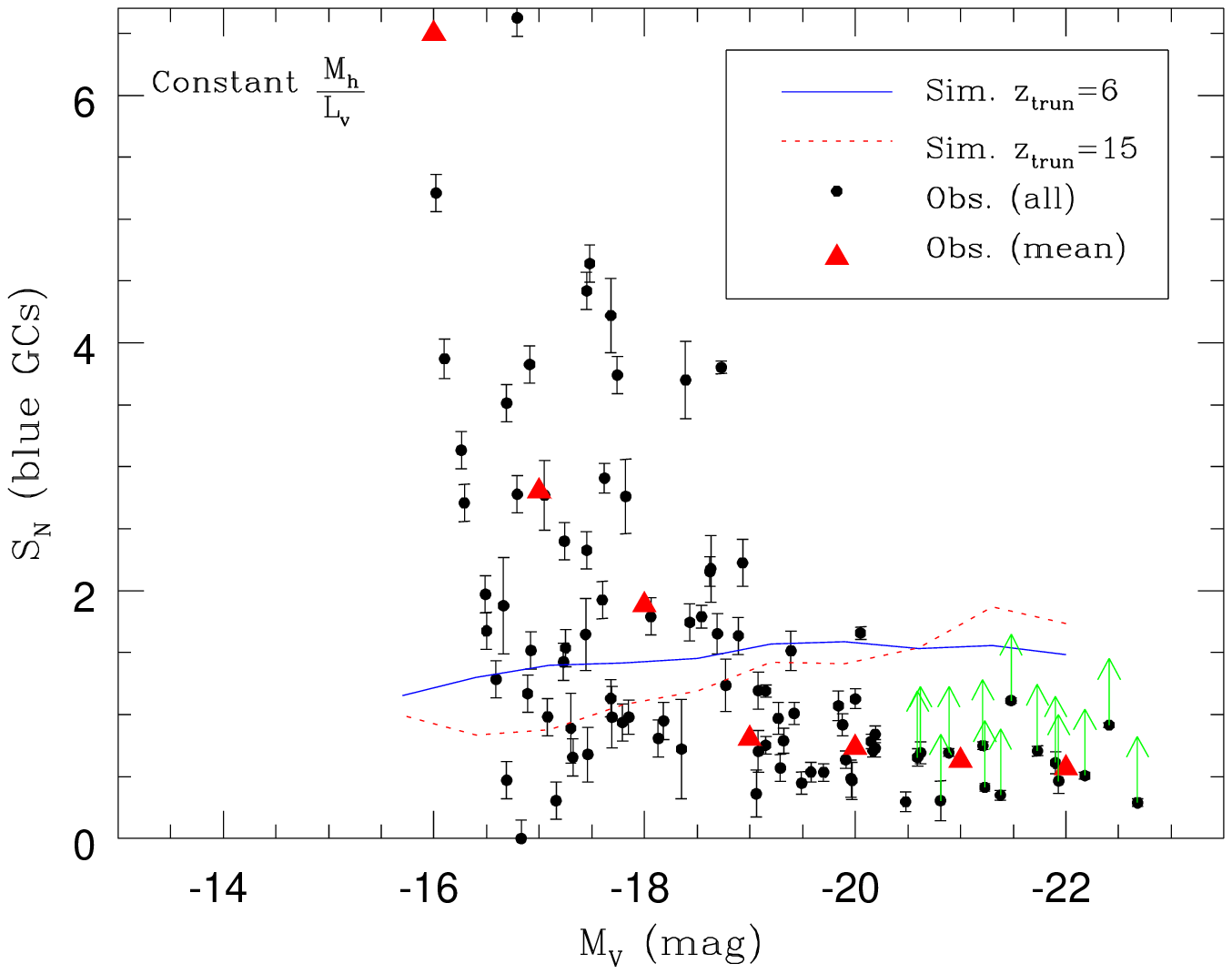}
\figcaption{
The same as Figure 3 but for the models with constant $M_{\rm h}/L_{\rm V}$
(=10).
The shapes of the derived $S_{\rm N}-M_{\rm V}$ relation 
can be slightly different from those of the $T_{\rm N}-M_{\rm h}$ one
shown in Figure 2, though overall shapes are quite similar
between the two relations for $M_{\rm h}>8.5 \times 10^8 {\rm M}_{\odot}$
(corresponding to  $M_{\rm V}<-15.0$ mag).
The origin of the slight differences is associated with
the difference in the binning process of simulation data 
between Figures 2 and 4.
\label{fig-4}}


\begin{thebibliography}{99}



\bibitem[]{576} Beasley, M.~A., Baugh, C.~M., Forbes, D.~A., Sharples, R.~M.,
Frenk, C.~S.\ 2002, MNRAS, 333,


\bibitem[]{580} Bekki,~K. 2005, ApJ, 626, L93


\bibitem[]{583} Bekki,~K., Chiba,~M. 2005, ApJ, 625, L107


\bibitem[]{589} 
Bekki, K., Forbes, D. A., Beasley, M. A., Couch, W. J.
2003, MNRAS, 344, 1334



\bibitem[]{598}
Benson, A. J., Nusser, A., Sugiyama, N., 
Lacey, C. G. 2001, \mnras, 320, 153


\bibitem[]{603} Bertschinger,~E. 1995, astro-ph/9506070 

\bibitem[]{605} Bertschinger,~E. 2001, \apjs, 137, 1





\bibitem[]{622} 
Durrell, P. R., Harris, W. E., Geisler, D., \& Pudritz, R.
1996, AJ, 112, 972


\bibitem[]{628} 
Forbes, D. A. 2005, ApJL, 635, 137

\bibitem[]{631} 
Harris, W. E. 1991, ARA\&A, 29, 543

\bibitem[]{634} 
Kaiser, N. 1984, ApJL, 284, 9



\bibitem[]{640}
Marinoni, C., Hudson, M. 2002, ApJ, 569, 101
	

\bibitem[]{646} 
Moore, B., et al. 2005, astro-ph/0510370

\bibitem[]{649} 
Navarro, J. F., Frenk, C. S., \& White, S. D. M. 1996, ApJ, 462, 563 (NFW)

\bibitem[]{652} 
Peng, E., et al. 2006, ApJ, 639, 95 

\bibitem[]{653} 
Rhode, K. L., Zepf, S. E.,  \& Santos, M. R. 2005, \apjl, 630, 21

\bibitem[]{656} Santos, M. R. 2003, 
in Extragalactic Globular Cluster Systems, Proceedings 
of the ESO Workshop,  p. 348

\bibitem[]{660} 
West, M. J. 1993, MNRAS, 265, 755

\bibitem[]{663} 
Wyithe, S. \& Loeb, A. 2004, Nature, 432, 194

\bibitem[]{666} 
Yahagi, H. 2005, PASJ, 57, 779  

\bibitem[]{669} 
Yahagi, H., \&  Yoshii, Y. 2001, \apj, 558, 463


\bibitem[]{673} 
Yahagi, H., Bekki, K. 2005,  \mnras, 364, L86 

\bibitem[]{676} 
Zepf, E., Ashman, K. M. 1993, MNRAS, 264, 611


\end{thebibliography}
\end{document}